\documentclass[12pt]{article}
\usepackage{graphicx}

\newenvironment{sciabstract}{%
\begin{quote} 
}
{\end{quote}}

\newcounter{lastnote}

\title{\bf \Large Universal collapse of the viscosity of supercooled fluids} 

\author{N. B. Weingartner$^{\dagger~1}$,  C. Pueblo${}^1$, F. S. Nogueira,${}^{2,3}$, K. F. Kelton${}^1$, and Z. Nussinov${}^{1\ast}$
\\
\\
\normalsize{\em $^{1}$Department of Physics and Institute of Materials Science and Engineering}\\
\normalsize{\em Washington University}\\
\normalsize{\em St. Louis, Missouri 63130, USA}\\
\normalsize{\em $^{2}$ Institute for Theoretical Solid State Physics,}\\
\normalsize{\em IFW Dresden, PF 270116, 01171 Dresden, Germany}\\
\normalsize{\em $^{3}$ Institut fur Theoretische Physik III}\\
\normalsize{\em Ruhr-Universitat Bochum, Universitatsstrasse 150}\\
\normalsize{\em DE-44801 Bochum, Germany}
\\
\normalsize{$^\ast$
Corresponding author:
{\tt zohar@wuphys.wustl.edu}}
}

\date{}

\begin{document} 


\baselineskip18pt


\maketitle


\begin{sciabstract}
{\bf{All liquids in nature can be supercooled to form a glass. Surprisingly, although this phenomenon has been employed for millennia  \cite{bib:tech1,bib:tech2,bib:tech3,bib:tech4,bib:tech5,bib:tech6}, it still remains ill-understood. Perhaps the most puzzling feature of supercooled liquids is the dramatic increase in their viscosity as the temperature ($T$) is lowered. This precipitous rise has long posed a fundamental theoretical challenge. Numerous approaches currently attempt to explain this phenomenon. When present, data collapse points to an underlying simplicity in various branches of science. In this Letter, we report on a 16 decade data collapse of the viscosity of 45 different liquids of all known types. 
Specifically, the viscosity of supercooled liquids scaled by their value at their respective equilibrium melting temperature ($\eta(T)/\eta(T_{melt}))$ is, 
for all temperatures $T<T_{melt}$, a universal function of $(T_{melt} - T)/(B T)$ where $B$ is a constant that does not change significantly from one liquid to another. This exceptionally plain behavior hints at {\it a link between glassy dynamics and the conventional equilibrium melting transition} in all known supercooled fluids.}}
\end{sciabstract}

\newpage
\noindent

A long standing endeavor of science is to understand one of the most prevalent and enigmatic states of matter- the ``glass'' 
\cite{bib:tech1,bib:paw,bib:rmp,bib:review}. {\it Any} system may be made glassy by supercooling (although the ease in which this can be achieved varies greatly). ``Supercooling'' refers to a rapid cooling of the liquid below its melting temperature so that crystallization cannot occur. As the temperature of supercooled liquids is depressed, they progressively become increasingly sluggish. At sufficiently low temperatures, when the viscosity ($\eta$) exceeds a threshold value of $10^{12}$ Pascal $\times$ second, the resulting system is termed a ``glass''. At temperatures below the ``glass transition temperature'' $T_{g}$ at which the above threshold is reached, the relaxation times are too long to be readily measured. Given their high viscosity, glasses exhibit a solid like rigidity. However, by comparison to ordered solids, glasses are highly amorphous. The ``transition'' of liquids into the glassy state is very odd. Most (non-glassy) systems exhibit marked changes in all measurable quantities as they undergo transitions from one phase to another. By stark contrast, while the dynamics of supercooled liquids may change dramatically (the viscosity and relaxation times of supercooled liquids may increase by many orders of magnitude en route to forming glass) \cite{bib:rmp,bib:review,bib:angell}, their static observables (such as structure and thermodynamic observables) display far milder changes.

\begin{figure}
\centering
\includegraphics[width=6.2in]{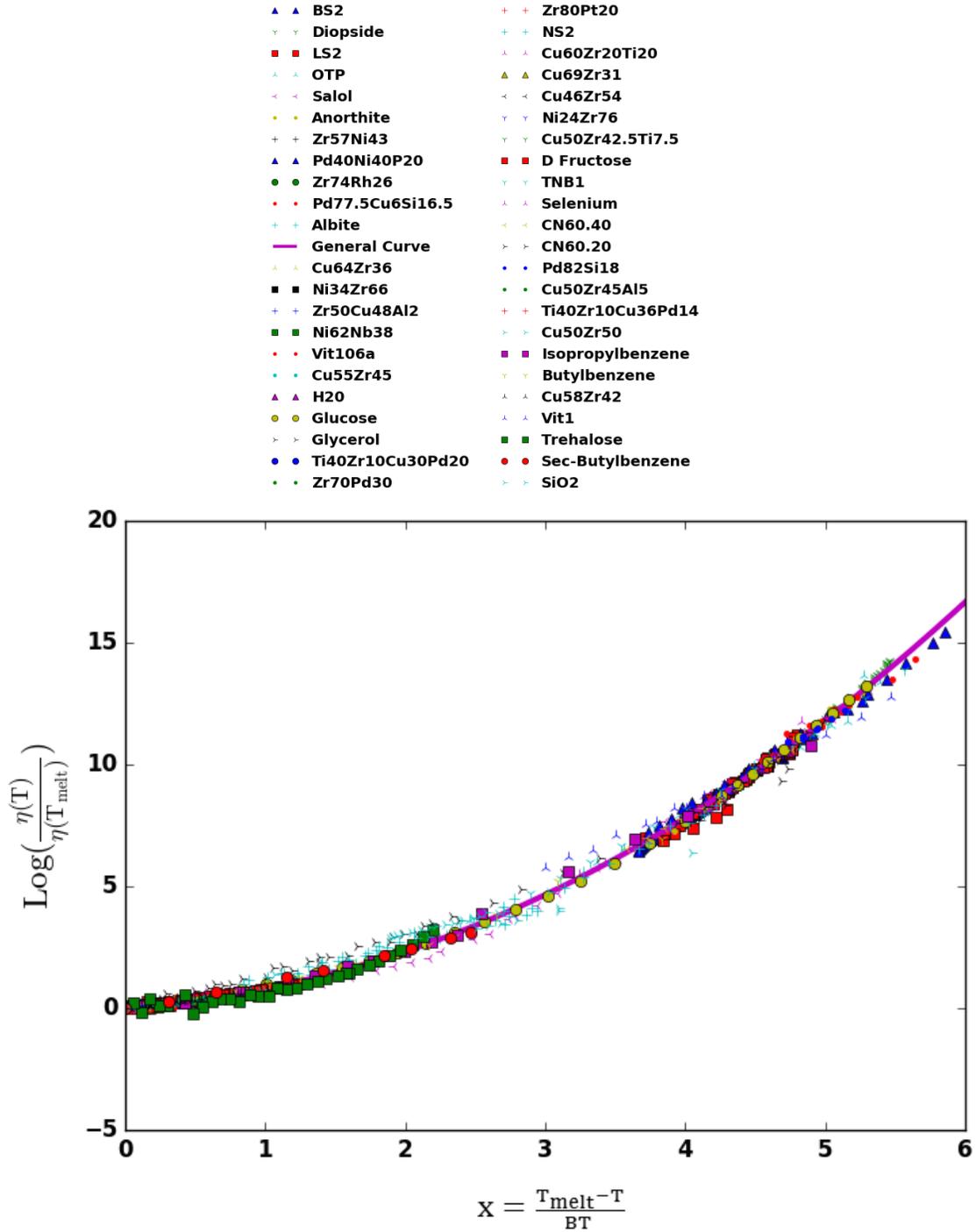}
\caption{\small \it 
{\em The scaling relation of Eq. (\ref{scaling}).  The``General Curve'' is given by Eq. (\ref{fz}).}}
\label{fig:collapse}
\end{figure}

The most common empirical (Vogel-Fulcher-Tamann-Hesse (VFTH)) \cite{bib:vft} fit for the viscosity of supercooled liquids is, by now, nearly a century old. According to this fit, the viscosity at a temperature $T$ is $\eta(T) = \eta_{0} e^{DT_{0}/(T-T_{0})}$ where $T_{0}, D$ and $\eta_{0}$ are liquid specific constants. This functional form asserts the existence of a singular temperature ($T_{0}$) at which the viscosity diverges; $T_{0}$ has often been termed an ``ideal glass transition temperature''. Since viscosity data at temperatures below $T_{g}$ have been scarce, it has been hard to examine this putative divergence. To date, the VFTH and all other fits and suggested theories argue for special temperature dependences that are not associated with the simple equilibrium melting transition temperature, e.g.,  \cite{bib:rmp,bib:review,bib:ag1,bib:ag2,bib:gt5,bib:CG,bib:mct1,bib:mct2,bib:mct3,bib:KKZNT1,bib:KKZNT2,bib:KKZNT3,bib:BENK1,bib:BENK2,bib:Blodgett,bib:MYEGA}. In essence, the existence of the glassy state has always been ascribed to particular processes that are notably different from those present in the standard {\it equilibrium melting (or freezing) transition}. A recent theoretical approach to the glass transition \cite{bib:DEH1} prompted us to critically revisit this deeply rooted dogma and to explore a simple possibility- that of a universal collapse of the viscosity data in which the sole important temperature is that of the standard equilibrium melting transition ($T_{melt}$). Here, $T_{melt}$ is the temperature at which slow cooling leads to the standard liquid to solid transition (or, more precisely, the ``liquidus'' temperature above which the system is a fluid). That is, we ask whether the glass transition might be simply related to the standard equilibrium melting transition without the need to invoke, from the outset, assumptions about conjectured novel temperatures (such as $T_{0}$), appealing to constrained dynamics/activation processes in a complex energy landscape, or invoke other numerous considerations. Towards this end, we analyzed 45 different supercooled fluids of all known types (including organic, silicate, metallic, and chalcogenide glassformers) to see whether the temperature dependence of the viscosity is dominated by the melting temperature. Specifically, we tested if a collapse of the form
\begin{eqnarray}
\frac{\eta(T)}{\eta(T_{m})} = F(\frac{T_{melt}-T}{B T})
\label{scaling}
\end{eqnarray}
occurs. Here, $B$ is a dimensionless material dependent constant and $F$ is a universal function. As seen in Fig. \ref{fig:collapse} , this relation is indeed satisfied. The values of $B$ do not vary widely across different glassformers (see Tables \ref{table} and \ref{table+} in the Supporting Information). Much further detail and analysis (at temperatures both below and above the melting transition temperature) appear in \cite{bib:DEH2}. The theory of \cite{bib:DEH1} predicts Eq. (\ref{scaling}) with a particularly succinct functional form,
\begin{eqnarray}
F(x)= \frac{1}{{\rm erfc}(x)}.
\label{fz}
\end{eqnarray} 
This prediction, highlighted in the collapse of Fig. \ref{scaling}, conforms well to the data.  
An assumption was made in deriving Eq. (\ref{fz}) that we will briefly comment on below and is elaborated on in great detail in \cite{bib:DEH1}.
Regardless of theoretical bias, the raw experimental data illustrate that Eq. (\ref{scaling}) holds very well 
(with a function that is equal to or is very close to Eq. (\ref{fz}) in the broad range of $x \equiv \frac{T_{melt}-T}{B T}$ values).


In conclusion, we demonstrated that the conventional equilibrium transition temperature $T_{melt}$ governs the dynamics of a supercooled fluid at temperatures $T<T_{melt}$. This finding lies contrary to approaches in which numerous other special temperatures have been suggested to dominate  the phenomenology of the glass transition. The work of \cite{bib:DEH1} first proposed Eq. (\ref{scaling}) by relying on the observation that the same many body Hamiltonian governs both the equilibrium liquid as well as the supercooled liquid. The core observation of \cite{bib:DEH1} was that the spectral problem posed by this single Hamiltonian has singularities only at the energy densities associated with the equilibrium melting transition (this is the only transition that occurs as the system's energy density (or, equivalently, the temperature) is lowered). This, in turn, implied that the only singularity governing the behavior of supercooled liquids should also be associated with these energies or, equivalently, with the equilibrium melting temperature. A (scale free) Gaussian distribution
in the eigenstate decomposition (in which the temperature $T$ was the only dimensional parameter governing the width of this distribution) led to Eq. (\ref{fz}). Albeit suggestive, our results should not be taken as a proof of this theory. Rather, we wish to underscore that the universal collapse of Fig. (\ref{fig:collapse}) is, on its own, a feature that underlies all glassformers. Our finding constitutes a challenge to the common lore that the standard equilibrium melting temperature does not play a central role in determining the dynamics of supercooled liquids. 

{\bf Acknowledgements:}  NW and ZN were supported by the NSF DMR-1411229. ZN thanks the Feinberg foundation visiting faculty program at Weizmann Institute. CP and KFK were supported by the NSF DMR 15-06553 This work was completed while ZN was at the Aspen Center for
Physics, which is supported by the National Science Foundation grant PHY-1066293.



{\small

\newpage

{\bf{Supporting Information}} 

In the tables that follow, the parameter $B$ in the collapse of  Eq. \ref{scaling} and Fig. \ref{fig:collapse} is provided. 
Further detail concerning the analysis leading to these values and many further aspects (both empirical and theoretical) appear in \cite{bib:DEH2}. 
  
\begin{table*}[t]
\centering
\caption{Values of Relevant Parameters for all liquids studied}
\begin{tabular}{*{4}{@{\hskip 0.4in}c@{\hskip 0.4in}}} 
\emph{Composition} \quad & \emph{}$B$ & \emph{$T_{melt}$ [K]} & \emph{$\eta(T_{melt})$ [Pa*s]} \\
BS2 & 0.157129 & 1699 & 5.570596   \\
Diopside & 0.134328 & 1664 & 1.5068     \\
LS2 & 0.170384 & 1307 & 22.198     \\
OTP & 0.069685 & 329.35 & 0.02954    \\
Salol & 0.087192  & 315 & 0.008884   \\
Anorthite & 0.131345 & 1823 & 39.81072   \\
$Zr_{57}Ni_{43}$ & 0.234171 & 1450 & 0.01564    \\
$Pd_{40}Ni_{40}P_{20}$ & 0.154701 & 1030 & 0.030197   \\
$Zr_{74}Rh_{26}$ & 0.187851 & 1350 & 0.03643    \\
$Pd_{77.5}Cu_6Si_{16.5}$ &  0.124879 & 1058 & 0.0446     \\
Albite & 0.103344 & 1393 & 24154952.8 \\
$Cu_{64}Zr_{36}$ &  0.142960 & 1230 & 0.021      \\
$Ni_{34}Zr_{66}$ & 0.209359 & 1283 & 0.0269     \\
$Zr_{50}Cu_{48}Al_{2}$ &0.167270 & 1220 & 0.0233     \\
$Ni_{62}Nb_{38}$ & 0.109488  & 1483 & 0.042      \\
Vit106a &0.133724  & 1125 & 0.131      \\
$Cu_{55}Zr_{45}$ & 0.144521 & 1193 & 0.0266     \\
$H_2O$ & 0.133069 & 273.15 & 0.001794   \\
Glucose & 0.079455 & 419 & 0.53       \\
Glycerol &  0.108834 & 290.9 & 1.9953     \\
$Ti_{40}Zr_{10}Cu_{30}Pd_{20}$ &0.185389  & 1279.226 & 0.01652    \\
$Zr_{70}Pd_{30}$ &  0.21073 & 1350.789 & 0.02288    \\
$Zr_{80}Pt_{20}$ & 0.169362 & 1363.789 & 0.04805    \\
NS2 & 0.134626 & 1147 & 992.274716 \\
$Cu_{60}Zr_{20}Ti_{20}$ & 0.103380 & 1125.409 & 0.04516    \\
$Cu_{69}Zr_{31}$         & 0.157480  & 1313     & 0.01155    \\
$Cu_{46}Zr_{54}$         &  0.156955 & 1198     & 0.02044535 \\
\hline       
\end{tabular}
\label{table}
\end{table*}

\begin{table*}[t] 
\centering
\caption{Values of Relevant Parameters for all liquids studied (continued)}
\begin{tabular}{*{4}{@{\hskip 0.4in}c@{\hskip 0.4in}}} 
\emph{Composition} \quad & \emph{}$B$ & \emph{$T_{melt}$ [K]} & \emph{$\eta(T_{melt})$ [Pa*s]} \\
$Ni_{24}Zr_{76}$         & 0.244979 & 1233     & 0.02625234 \\
$Cu_{50}Zr_{42.5}Ti_{7.5}$  & 0.148249 & 1152     & 0.0268     \\
D Fructose       &  0.050124 & 418      & 7.31553376 \\
TNB1             &  0.07567 & 472      & 0.03999447 \\
Selenium         & 0.130819  & 494      & 2.9512     \\
CN60.40          & 0.149085  & 1170     & 186.2087   \\
CN60.20          & 0.161171  & 1450     & 12.5887052 \\
$Pd_{82}Si_{18}$         &  0.137623 & 1071     & 0.03615283 \\
$Cu_{50}Zr_{45}Al_{5}$     & 0.118631   & 1173     & 0.03797    \\
$Ti_{40}Zr_{10}Cu_{36}Pd_{14}$ &  0.137753 & 1185     & 0.0256     \\
$Cu_{50}Zr_{50}$       & 0.166699 & 1226     & 0.02162    \\
Isopropylbenzene &   0.073845 & 177      & 0.086      \\
ButylBenzene     & 0.085066  & 185      & 0.0992     \\
$Cu_{58}Zr_{42}$         &  0.131969 & 1199     & 0.02526    \\
Vit 1 & 0.111185 & 937 & 36.59823 \\
Trehalose & 0.071056  & 473 & 2.71828 \\
Sec-Butylbenzene & 0.080088 & 190.3 & 0.071 \\
$SiO_2$ & 0.090948 & 1873 & 1.196x$10^{8}$ \\
\hline       
\end{tabular}
\label{table+}
\end{table*}

\end{document}